\documentclass[12pt,a4paper]{article}

\usepackage{amsmath}
\usepackage{amssymb}
\usepackage{float}
\usepackage{tensor}
\usepackage{amsfonts}
\usepackage{dsfont}
\usepackage{braket}
\usepackage{bm}
\usepackage{cite}
\usepackage{setspace}
\usepackage{mathrsfs}
\usepackage{amsthm}
\usepackage{enumerate,graphicx}
\usepackage[colorlinks,linktocpage]{hyperref}
\usepackage{xcolor}
\definecolor{rood}{RGB}{230,36,0}
\definecolor{groen}{RGB}{55,210,0}
\hypersetup{citecolor=groen, linkcolor=rood}
\usepackage{placeins}
\usepackage{titlesec}
\usepackage[small,bf,hang]{caption}
\usepackage[left=2.5cm,right=2.5cm,top=2.6cm,bottom=2.5cm]{geometry}
\usepackage{subcaption}
\usepackage[T1]{fontenc}

\DeclareMathAlphabet{\mathpzc}{OT1}{pzc}{m}{it}
\DeclareMathOperator{\tr}{tr}

\newcommand{\vt}[1]{\ensuremath{\boldsymbol{#1}}}

\begin{document}
\renewcommand{\baselinestretch}{1.10}\normalsize

\title{Metrological measures of non-classical correlations}
\author{Pieter Bogaert and Davide Girolami}


\maketitle
 
\pagenumbering{arabic}

\noindent{\it Affiliation: Department of Atomic and Laser Physics, University of Oxford, Parks Road, OX1 3PU, United Kingdom}\\
{\it Correspondence: davegirolami@gmail.com}
\section{Introduction}

In this work, we will review studies showing that non-classical, discord-like correlations do not necessarily describe a statistical dependence between measurements performed by non-communicating parties. We will explain how they yield the impossibility of global observers to obtain full knowledge of local properties of quantum systems. This apparently detrimental feature translates, on the other hand, in an increased capability of an observer to acquire information about a quantum perturbation by establishing correlations between its probe and an unchanged ancillary system. The phenomenon is undoubtedly not explicable by classical physics, being a direct consequence of quantum complementarity. We will present our arguments by following a two-step line of thinking.\\
First, we will point out that quantum coherence manifests in the intrinsic quantum randomness of measurement outcomes  (Sec \ref{uncertainty}). Genuinely quantum uncertainty differs from classical randomness. We will explain how to discriminate between them and quantify the quantum uncertainty from experimental data. Non-classical correlations in a bipartite system will be defined as the degree of irreducible coherence, i.e. quantum uncertainty or randomness, experienced when measuring local observables. The result links a local property as quantum uncertainty to a global feature as non-classical correlations. The proof is given by showing that a quantity called Local Quantum Uncertainty, which quantifies the minimum local quantum randomness in a bipartite state, satisfies the very same properties obeyed by entropic measures of discord-like correlations (Sec. \ref{lqu}). \\
Then, we will discuss the interplay between quantum-induced uncertainty and supraclassical measurement precision (Sec. \ref{metrology}).  A measurement can be thought of as an information processing task where knowledge encoded in a physical systems is transmitted to an apparatus. Specifically, a measurement requires a preliminary step in which the probe is prepared in an input configuration. In a second stage, the information we want to access is imprinted in the probe state through quantum dynamics. The final part is the 
information decoding by collection and statistical analysis of the data. We will focus here on the first step, i.e. input state preparation. 
Arguably, the probe state has to be sensitive to the perturbation. We will explain why quantum systems displaying non-classical correlations are intrinsically more sensitive probes. The key observation is that quantum uncertainty entails sensitivity to quantum dynamics.  Consequently, non-classical correlations guarantee non-vanishing sensitivity to local quantum perturbations.   We will explain how the concept of Interferometric Power captures genuinely quantum sensitivity in a standard measurement setting, and how this leads to non-classical performances for phase estimation (Sec. \ref{intpower}). Remarkably, the minimum precision for local measurements will be shown to be a measure of non-classical correlations. A third interesting correlation quantifier, the Discriminating Strength (Sec. \ref{ds}), will be shown to evaluate the worst case precision in another important metrological task, state discrimination.

It is our hope to highlight the main merit, in our opinion, of the metrological approach to characterising non-classical correlations.  That is, giving a physical meaning to an information-theoretic construction and providing an operational interpretation which goes truly beyond the original one\cite{conf}.  Quantum discord is a concept developed to study environmentally induced decoherence \cite{OZ}, and the limit to information transmission established by classical correlations \cite{HV}. Other concurrent studies characterised non-classical correlations in the context of quantum Shannon information theory \cite{modirev}. While Quantum Mechanics is somehow a theory of information itself, we owe its postulates and structure to key experimental observations of low-energy light and atomic structure in the beginning of the 20th century \cite{dirac}. For example, the somehow elusive concept of Entanglement was originally discussed by means of carefully designed thought experiments. It is therefore reassuring to make real  the concept of non-classical correlations by linking it with observable experimental effects. \\

\section{Local Quantum Uncertainty}\label{uncertainty} 

\subsection{Quantum uncertainty}
\begin{figure}
 \centering
 \includegraphics[height=6.6cm,width=10cm]
 {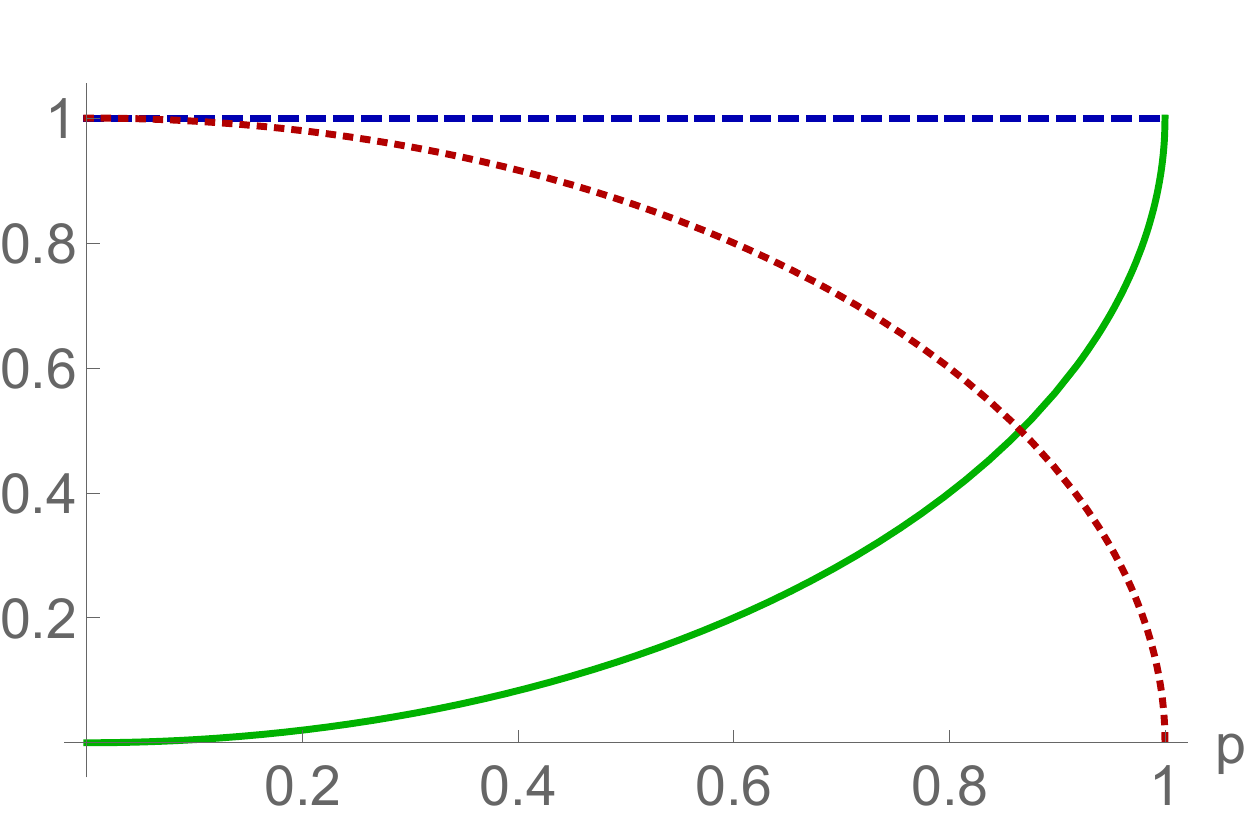}
\caption{Quantum uncertainty disclosed. We calculate the uncertainty on the measurement outcome of
 the observable $\sigma_z=\ket{0}\bra{0}-\ket{1}\bra{1}$
in the state $\rho =
(1 - p)\mathbb{I}_2/2+ p \ket{\phi}\bra{\phi}, \ket{\phi}= 1/\sqrt 2(\ket{0}+\ket 1), p\in[0, 1]$. The blue dashed
line is the variance, the green blue continuous curve is the skew information. The red
dotted curve depicts the difference between the two quantities, being a heuristic mixedness quantifier. As expected by a measure of quantum uncertainty, the skew information monotonically increases with the purity parameter $p$.}\label{fig1}
\end{figure}

 Quantum Mechanics predicts the existence of coherent superpositions of quantum states \cite{dirac}. The first experimental evidence which suggested such a possibility was the wave-like probability distribution of measurement outcomes observed in low-energy optical experiments \cite{taylor}. The intuition linking coherence and non-classical outcome statistics can be formalised. By focusing on finite-dimensional quantum systems, let us suppose to measure the observable  being represented by a non-degenerate Hermitian operator with spectral decomposition $O=\sum_i o_i \ket{i_o}\bra{i_o}$. The information  about $O$ in a state represented by a density matrix $\rho, \tr[\rho]=1,\rho=\rho^{\dagger}, \rho\geq0,$ can be quantified by the state change due to the measurement (without postselection) of $O$. For our purposes, we focus on the von Neumann measurement model $\rho\rightarrow \rho'= \sum_i  \ket{i_o}\bra{i_o}\rho\ket{i_o}\bra{i_o}$  \cite{vn}. If and only if the state and the observable commute, there is no change in the state, $\rho=\rho'$.  This is easily proven to happen if and only if the state is an eigenstate or a mixture of eigenstates of the observable, taking the form $\rho_O=\sum_i p_i \ket{i_o}\bra{i_o}$. That is, if and only if the state is incoherent in the observable eigenbasis,  the measurement output statistics will be classical. Without coherence, the measurement uncertainty is only due to incomplete knowledge of the system state, which is a classical error source. In the more general case of states displaying coherence, the contribution to the measurement uncertainty is then twofold. Apart from the classical randomness, there is an additional quantum component, which manifests itself in the interference pattern of the outcome statistics.  \\
Let us now quantify quantum uncertainty. The first quantity that has in many ways become almost synonymous with uncertainty, at least in undergraduate Physics textbooks, is the variance $V(\rho,O)=\tr[\rho O^2]-(\tr[\rho O])^2$. The variance enjoys both a simple expression and a close tie to experimental practice. However, for mixed states the variance includes a contribution of classical uncertainty due to the mixedness of the state. It is easy to see that the variance does not vanish even if $\rho$ and $O$ commute, apart from the case in which the state is an observable eigenstate. The variance is therefore not suitable to quantify quantum uncertainty.
A way to solve the issue is to formally split the variance, which captures the total measurement uncertainty, into quantum and classical contributions: $V=V_q+V_c$ \cite{luo}. Note that the idea can be extended to entropic uncertainty quantifiers \cite{herbut}. A good measure of quantum uncertainty $V_q$  should be zero if and only if $\rho$ and $O$ commute. Yet, an arbitrary norm of their commutator is not the finest choice. Additionally, a measure of quantum uncertainty should be convex, i.e. non-increasing under classical mixing, as this   only generates classical uncertainty, $V_q\left(\sum_ip_i\rho_i,O\right)\leq\sum_ip_i V_q\left(\rho_i,O\right)$.
 A suitable candidate is  the (Wigner-Yanase) skew information \cite{skew}, given by
\begin{equation}\label{skew}
\mathcal{I}(\rho,O):=-\frac{1}{2}\tr [[\rho^{1/2},O]^2].\end{equation}
The skew information is upper bounded by the variance, being equal to it for pure states: $\mathcal{I}(\rho,O)\leq V(\rho,O)$. This can be shown as follows \cite{luo}. By defining $O_0=O-\tr[\rho O]$, one has $V(\rho,O)=\tr[\rho O_0^2]$ and $\mathcal{I}(\rho,O)=\tr[\rho O_0^2]-\tr[\rho^{1/2}O_0\rho^{1/2}O_0]=V(\rho,O)-\tr[\rho^{1/2}O_0\rho^{1/2}O_0].$
It is easy to see that the second term is non-negative as it equals $\tr[(\rho^{1/4}O_0\rho^{1/4})(\rho^{1/4}O_0\rho^{1/4})]$ and noting that $\rho^{1/4}O_0\rho^{1/4}$ is self-adjoint. Whilst being just one of the potential choices, the skew information is a consistent yet sufficiently manageable measure of quantum uncertainty. We illustrate the interplay between classical and quantum uncertainty by a simple example presented in Fig.\ref{fig1}.

\subsection{Discord triggers local quantum uncertainty}\label{lqu}
The Heisenberg uncertainty principle states that complementary properties of quantum systems cannot be measured with arbitrary precision, in the sense that, regardless of our experimental ability, the product of the experimental uncertainties about their values in a given state is lower bounded by the size of their commutator \cite{heis}, which thus captures such ineludible quantum randomness. 
In the original form of the uncertainty relations, the non-commutativity between observables captures such  ineludible quantum randomness. However, it may seem that any single physical quantity, such as one spin or position component, could be measured with arbitrary precision. We are going to show that this is not true in general. We identify the truly quantum uncertainty of the measurement, and, not surprisingly, quantify it by a measure of state-observable non-commutativity.  Zero quantum uncertainty implies that the measurement performed by a flawless experimental implementation, i.e. whenever there is not even classical uncertainty, has a deterministic outcome. Yet, a non-negotiable intrinsic quantum uncertainty on single observable measurements appears whenever the system of interest  shares non-classical correlations.\\
Let us examine the quantum uncertainty in local quantum measurements on a bipartite system. For example, let us consider a two-qubit system prepared in a maximally entangled state $\ket{\varphi}_{AB}=(\ket{00}+\ket{11})/\sqrt{2}$. It is immediate to observe that this is an eigenstate of the global observable $\sigma_z\otimes\sigma_z$, which means that there is no quantum uncertainty when measuring that observable. Any {\it local} spin measurement, however, will have intrinsic uncertainty. The only vector $\vt{n}$ for which $\vt{n}\cdot\vt{\sigma}_A\otimes\mathbb{I}_B\ket{\varphi}_{AB}=k\ket{\varphi}_{AB}$, $k\in\mathds{R}$, where $\vt{\sigma}$ are the Pauli matrices, is indeed $\vt{n}=\vt{0}$. More generally, only product states (e.g. $\ket{11}$) can be eigenstates of local observables.\\
By extending the argument to mixed states, it is clear that one does not want to associate quantum uncertainty to state mixedness (which quantifies the incomplete knowledge about the state).  Given a local complete measurement, we still require that performing the measurement leaves the mixed state $\rho_{AB}$  invariant if and only if it commutes with the observable. Supposing without loss of generality that the measurement is performed on $A$, this means that it must be possible to express the state in the following form:
\begin{equation}\rho_{AB}=\sum_ip_i\ket{i}\bra{i}_A\otimes\sigma_B^i,\end{equation}
where the elements $\{ \ket{i}\}$ form an orthonormal basis.  Such density matrices are called classical-quantum (CQ) states, and they are precisely the states with zero quantum discord \cite{modirev}. Therefore, non-classical correlations imply local quantum uncertainty. In other words, for any CQ state there is {\it at least} one local measurement which does not alter it, while for  other states quantum uncertainty always appears.  
However, the interplay between local randomness and non-local quantum effects turns out to be deeper. The minimum  quantum uncertainty on local measurements is a quantifier of non-classical correlations. To prove that, let us quantify the quantum uncertainty of an  observable $O_A$ in a state $\rho_{AB}$ by the skew information ${\cal I}(\rho_{AB},O_A\otimes\mathbb{I}_B)$. By reminding the definition in Eq. \ref{skew}, we note that the quantity depends on the state  and the observable, while non-classical correlations are a property of the state only. It is sensible to introduce the Local Quantum Uncertainty (LQU) \cite{lqu}, defined as the skew information between the state and a local observable, minimised over all local observables. To be more precise, let us define the set of local observables $\{K^\Lambda_A:=K^\Lambda_A\otimes\mathbb{I}_B\}$, where the $K^\Lambda_A$ are Hermitian operators with spectrum $\Lambda$, which we demand to be non-degenerate, as this would represent an additional classical uncertainty source. Thus, the LQU with respect to the subsystem $A$ is given by
\begin{equation}\mathcal{U}^\Lambda_A(\rho_{AB}):=\min_{K^\Lambda_A}\mathcal{I}(\rho_{AB},K_A^\Lambda),\end{equation}
with an optimisation over the previously defined set of local observables with non-degenerate spectrum $\Lambda$. We   rewrite them as $K^\Lambda_A=U_A\mathrm{diag}(\Lambda)U_A^\dagger, U_A \in SU(d),$ where $d$ is the dimension of subsystem $A$ and $\mathrm{diag}(\Lambda)$ is a diagonal matrix with the observable eigenvalues being the diagonal entries. The minimisation then runs over all possible unitary transformations $U_A$. The LQU is still dependent on the spectrum $\Lambda$, and this can be interpreted as fixing a ``ruler'' for the measurement. The  non-degeneracy condition ensures the quality of the ruler, namely that there exist states for which a measurement will be maximally informative (i.e. states which commute with the observable and hence do not exhibit quantum uncertainty for it). Any spectrum choice identifies a different measure of non-classical correlations. On the other hand, the LQU is by no means dependent on the measurement basis, as $U_A$ is varied over SU($d$).

\subsubsection{Local Quantum Uncertainty as a measure of non-classical correlations}\label{proofs}
We here review the proof that the LQU is a measure for non-classical correlations, i.e. it meets the criteria identifying discord-like quantifiers \cite{modirev}. We will always work with the LQU defined by measurements on $A$.
\begin{itemize}
 \item[1] The LQU is zero if and only if the state is CQ. If $\rho_{AB}$ is CQ, then one can pick a $K^\Lambda_A$ which is diagonal in the local basis of $A$, which means that the LQU vanishes. Conversely, if the LQU is zero, then there exists a local observable $K^\Lambda_A$ which is simultaneously diagonalisable with $\rho_{AB}$. Since $\Lambda$ is non-degenerate, this defines a basis on A which is unique up to phases (let us call it $\{\ket{k_i}\}$). An eigenvector basis for $K^\Lambda_A$ must then be of the form $\{\ket{k_i}_A\otimes\ket{\varphi_{ij}}_{B}\}$, and the state must therefore be of the form $\rho_{AB}=\sum_{ij}p_{ij}\ket{k_i}\bra{k_i}_A\otimes\ket{\varphi_{ij}}\bra{\varphi_{ij}}_B$, i.e. it must be CQ.
\item[2]The LQU is invariant under local unitary transformations.
A few algebra steps give
\begin{eqnarray}
\mathcal{U}^\Lambda_A((U_A\otimes U_B)\rho_{AB}(U_A\otimes U_B)^\dagger)&=&\min_{K^\Lambda}\mathcal{I}((U_A\otimes U_B)\rho_{AB}(U_A\otimes U_B)^\dagger,K^\Lambda_A\otimes\mathbb{I}_B)\nonumber\\
&=&\min_{K^\Lambda}\mathcal{I}(\rho_{AB},(U_A\otimes U_B)^\dagger K^\Lambda_A\otimes\mathbb{I}_B (U_A\otimes U_B))\nonumber\\
&=&\min_{K^\Lambda}\mathcal{I}(\rho_{AB},(U_A^\dagger K^\Lambda_A U_A)\otimes \mathbb{I}_B)=\mathcal{U}^\Lambda_A(\rho_{AB}),
\end{eqnarray}
where the second and third lines follow from the definition of the skew information. The last equality holds because minimising over $K^\Lambda_A$ is equivalent to minimising over the observable $U_A^\dagger K^\Lambda_A U_A$.
\item[3] The LQU is contractive under completely positive trace-preserving (CPTP) maps on the non-measured subsystem $B$. The skew information is contractive under CPTP maps $\Phi_B$: $\mathcal{I}(\rho_{AB},K_A\otimes\mathbb{I}_B)\geq\mathcal{I}((\mathbb{I}_A\otimes\Phi_B)\rho_{AB},K_A\otimes\mathbb{I}_B)$. This can be easily proved by writing the CPTP map $\Phi_B$ in a Stinespring representation and noting that the skew information is contractive under partial trace: $\mathcal{I}(\sigma_{AB},X_A\otimes\mathbb{I}_B)\geq\mathcal{I}(\sigma_A,X_A)$. Let us suppose now that $\tilde{K}_A$ is the local observable minimising the skew information. The LQU takes the form
\begin{equation}\mathcal{U}^\Lambda_A(\rho_{AB})=\mathcal{I}(\rho_{AB},\tilde{K}_A\otimes\mathbb{I}_B)\geq\mathcal{I}((\mathbb{I}_A\otimes\Phi_B)\rho_{AB},\tilde{K}_A\otimes\mathbb{I}_B)\geq\mathcal{U}^\Lambda_A((\mathbb{I}_A\otimes\Phi_B)\rho_{AB}).\end{equation}

\item[4] The LQU reduces to an entanglement monotone for pure states. For the full proof of this property, we refer to \cite{lqu},  presenting here just a sketch of it. Given the contractivity and invariance under CPTP and unitary maps respectively, we only need to prove that the LQU cannot increase on average under local operations on $A$:
\begin{equation}\sum_ip_i\mathcal{U}^\Lambda_A(\ket{\phi_i}\bra{\phi_i}_{AB})\leq\mathcal{U}^\Lambda_A(\ket{\psi}\bra{\psi}_{AB}),
\end{equation}
where $\{p_i,\ket{\phi_i}\}$ is the output ensemble after a channel with Kraus operators $\{M_i\}$ is applied on $A$: $M_{i,A}\ket{\psi}_{AB}=\sqrt{p_i}\ket{\phi_i}_{AB}$. It is possible to prove two auxiliary lemmas. First, one can always assume $d_A\geq d_B$. Then, one shows that the LQU is not affected when measuring $B$ instead of $A$, where $\Lambda(K_B)$ is a subset of $\Lambda(K_A)$. Suppose that the minimum is achieved for $\tilde{K}^\Lambda_B$. Since the skew information is equal to the variance for pure states, and the latter is concave, one finally has
\begin{eqnarray}
\sum_i p_i\mathcal{U}^\Lambda_A(\ket{\phi_i}\bra{\phi_i}_{AB})&\leq&\sum_ip_i\min_{K^\Lambda_B}\mathcal{I}(\ket{\phi_i}\bra{\phi_i}_{AB},K^\Lambda_B)\leq\sum_ip_i\mathcal{I}(\ket{\phi_i}\bra{\phi_i}_{AB},\tilde{K}^\Lambda_B)\nonumber\\
&=&\sum_ip_iV(\ket{\phi_i}\bra{\phi_i}_{AB},\tilde{K}^\Lambda_B)\leq V\left(\sum_ip_i\ket{\phi_i}\bra{\phi_i}_{AB},\tilde{K}^\Lambda_B\right)\nonumber\\
&=&\sum_ip_i\bra{\phi_i}(\tilde{K}^\Lambda_B)^2\ket{\phi_i}_{AB}-\left(\sum_ip_i\bra{\phi_i}\tilde{K}^\Lambda_B\ket{\phi_i}_{AB}\right)^2\nonumber\\
&=&\sum_i\bra{\psi}M_i(\tilde{K}^\Lambda_B)^2M^\dagger_i\ket{\psi}_{AB}-\left(\sum_i\bra{\psi}M_i\tilde{K}^\Lambda_BM^\dagger_i\ket{\psi}_{AB}\right)^2\nonumber\\
&=&\bra{\psi}(\tilde{K}^\Lambda_B)^2\ket{\psi}_{AB}-\left(\bra{\psi}\tilde{K}^\Lambda_B\ket{\psi}_{AB}\right)^2\nonumber\\
&=&\mathcal{I}(\ket{\psi}\bra{\psi}_{AB},\tilde{K}^\Lambda_B)=\min_{K^\Lambda_A}\mathcal{I}(\ket{\psi}\bra{\psi}_{AB},K^\Lambda_A)\nonumber\\
&=&\mathcal{U}^\Lambda_A(\ket{\psi}\bra{\psi}_{AB}).\end{eqnarray}
\end{itemize}

\subsubsection{Restriction to $\mathds{C}^2\otimes\mathds{C}^d$}

We now consider the case where system $A$ is a qubit and $B$ a qudit, i.e. with states defined on an Hilbert space $\mathds{C}^2\otimes\mathds{C}^d$. A question that remains to be answered is in which way the LQU depends on the choice of non-degenerate spectrum $\Lambda$. It is straightforward to show that, since $A$ is a qubit, all $\Lambda$-dependent $\mathcal{U}^\Lambda(\rho_{AB})$ are equivalent up to a multiplicative factor. This is because a general local observable $K^\Lambda_A$ with non-degenerate spectrum $\Lambda=\{\lambda_1,\lambda_2\}$ can be parametrised as
\begin{equation}K_A^\Lambda=U_A\left(\frac{\lambda_1-\lambda_2}{2}\sigma_{zA}+\frac{\lambda_1+\lambda_2}{2}\mathbb{I}_A\right)U_A^\dagger=\frac{\lambda_1-\lambda_2}{2}\vt{n}\cdot\vt{\sigma}_A+\frac{\lambda_1+\lambda_2}{2}\mathbb{I}_A, \end{equation}
where $\vt{n}$ is a unit vector. From the definition of the skew information, it follows that $\mathcal{I}(\rho_{AB},K^\Lambda_A)=\frac{(\lambda_1-\lambda_2)^2}{4}\mathcal{I}(\rho_{AB},\vt{n}\cdot\vt{\sigma}_A)$. Therefore, for qubit-qudit systems the choice of the spectrum $\Lambda$ does not affect the quantification of non-classical correlations (we will therefore drop the $\Lambda$ superscript from here onwards), and without loss of generality, we assume the local observables to be of the form $K_A=\vt{n}\cdot\vt{\sigma}_A$.

Having simplified the form of the  observables over which we need to optimise (the minimisation runs over $\vt{n}$ now), we can write the LQU in the following fashion:
\begin{equation} \mathcal{U}_A(\rho_{AB})=1-\lambda_\mathrm{max}(W_{AB}),\label{eq:lqu_qubitqudit}\end{equation}
$\lambda_\mathrm{max}(W_{AB})$ being the maximum eigenvalue of the 3$\times$3 symmetric matrix $W$ with entries
\begin{equation}(W_{AB})_{ij}=\tr[\rho_{AB}^{1/2}(\sigma_{iA}\otimes\mathbb{I}_B)\rho_{AB}^{1/2}(\sigma_{jA}\otimes\mathbb{I}_B)],\end{equation}
where $i,j$ label the Pauli matrices. Finally, for pure states $\ket{\psi}\bra{\psi}_{AB}$, this further reduces to (twice) the  linear entropy of entanglement   
\begin{equation}\mathcal{U}_A(\ket{\psi}\bra{\psi}_{AB})=2(1-\tr[\rho_A^2])=1-(\sigma_0-\sigma_1)^2,\end{equation}
where we used the Schmidt coefficients $\rho_A=\sigma_1\ket{\psi_1}\bra{\psi_1}_A+\sigma_2\ket{\psi_2}\bra{\psi_2}_A$. We  observe  that with our choice of  observables $K_A$ the LQU equals one for pure, maximally entangled states.

\subsubsection{Geometric insight}
Finally, we  provide a geometric interpretation of the LQU in qubit-qudit states. The (squared) Hellinger distance \cite{bengtsson, luo2} between two states $\rho$ and $\sigma$ is defined as $D^2_H(\rho,\sigma)=(1/2)\tr[\rho^{1/2}-\sigma^{1/2}]^2=1-\tr[\rho^{1/2}\sigma^{1/2}]$. Since $K_A=\vt{n}\cdot\vt{\sigma}$ is a root-of-unity unitary,   for every function $f$ and any bipartite state one has $K_Af(\rho_{AB} )K_A=f(K_A\rho_{AB} K_A)$. Hence,  the skew information  takes the form
\begin{eqnarray}{\cal I}(\rho_{AB},K_A)&=&1-\tr[\rho_{AB}^{1/2}K_A\rho_{AB}^{1/2}K_A]=1-\tr[\rho_{AB}^{1/2}(K_A\rho_{AB} K_A)^{1/2}]\nonumber\\
&=&D^2_H(\rho_{AB} ,K_A\rho_{AB} K_A).
\end{eqnarray}
The LQU then represents the minimum distance between the state before and after a local root-of-unity unitary operation is applied.

\section{Interferometric Power and Discriminating Strength}\label{metrology}

\subsection{Quantum Metrology}
We discussed a measure of discord-like correlations, the LQU, linked to the uncertainty in a given measurement. Perhaps surprisingly, in this section we will show that non-classical correlations yield measurement precision! We will explain how the two apparently contradictory viewpoints  are consistently related to each other in the context of quantum metrology, which we briefly introduce here. \\
Metrology is the study of measurement strategies and tools. The term can be used in a variety of contexts related to measurements, for example to denote the establishment of units of measurement, or the technological application of measurement instruments and related issues such as calibration. For our purposes, however, metrology denotes the study of parameter estimation schemes and the strategies to reach the highest possible precision in them. Many of the concepts in metrology were first defined for classical systems, but we will only discuss the ones which are useful for the extension to the quantum realm. For  enjoyable reviews on quantum metrology, we refer the Reader to Refs. \cite{toth,metrorev}. It is indeed possible to take advantage of quantumness to increase the precision of measurement schemes. The reason is that quantum systems are more sensitive probes in a number of situations. Quantum metrology is the research line that studies which properties of quantum systems  are responsible for this.  Results in quantum metrology have a wide applicability in  optical interferometry, atomic spectroscopy, and even gravitometry. 
A metrology task usually consists of three steps. First, the preparation of a probe in an input state. Second, an interaction or perturbation of the probe, which encodes information in it.  Third, a measurement on the probe followed by data analysis. Here we focus on the first step, and we investigate how non-classical correlations in the input help in two important metrology protocols: interferometric phase estimation and state discrimination.

\subsection{Quantum phase estimation}
\begin{figure}
\centering
 \includegraphics[height=6.6cm,width=10cm]{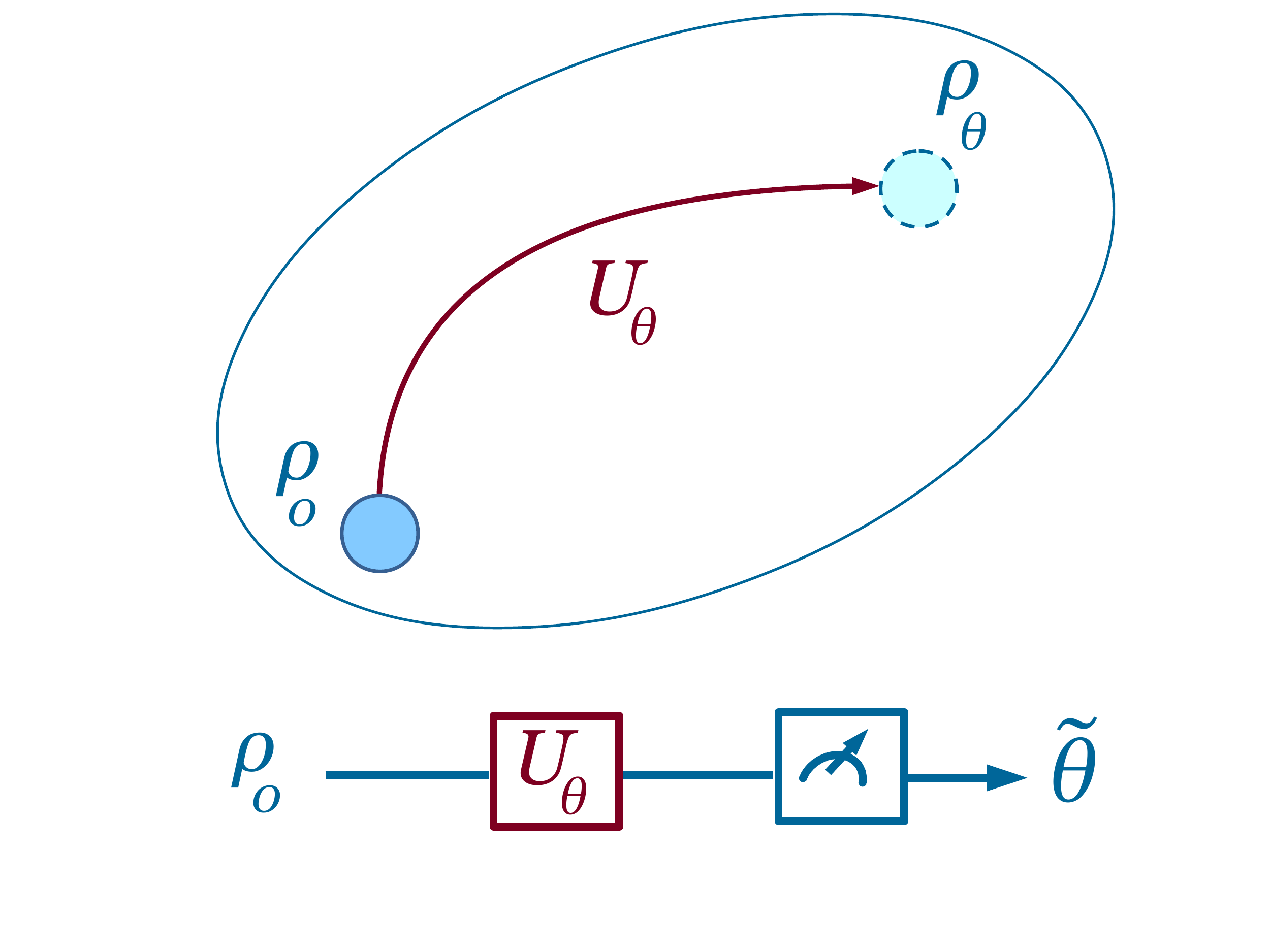}
\caption{Quantum phase estimation. A system initialized in the state $\rho_0$ is perturbed through a unitary transformation $U_{\theta}$. A measurement and statistical processing of outcomes give an estimated value $\tilde{\theta}$ of the phase shift. The perturbation can be represented both as a geometric path in the parametrized space of quantum states, where $\theta$ is a coordinate (top), or as a logic transformation by applying a unitary gate (bottom). The resource is found to be the speed of evolution of the state during the phase shift, as quantified by the quantum Fisher information.}
\label{fig2}
\end{figure}

We focus on the important metrology primitive of  parameter estimation \cite{helstrom}.
  The goal is to assign a probability function $p_{\theta}(x)$ to the independent measurement outcomes  $x$ of a random variable $X$. 
The parameter $\theta$, which is unknown and unmeasurable, acts as a coordinate in the probability function space. The task is then to extract an observable estimator $\hat\theta(x)$ from the measurement outcomes, such that $p_{\hat\theta}(x)$ characterizes well the observed data.  We require the estimator to be unbiased, i.e. its average value does equal the real value of the parameter, $\int(\theta-\hat{\theta}(x))p_{\theta}(x)dx=0$.  The quality of the estimation can be then quantified by the variance of the estimator $\hat{\theta}$. \\
It is possible to establish a fundamental limit to parameter estimation. By employing the maximum likelihood method,  the best estimator $\hat{\theta}_{\text{best}}$ is defined as the one maximising the log-likelihood function $\max\limits_{\hat\theta}\ln l(\hat\theta|x)=\ln l(\hat{\theta}_{\text{best}}|x),  l(\hat\theta|x)\equiv p_{\hat\theta}(x)$, where the logarithm is just a convention. This means that $p_{ \hat{\theta}_{\text{best}}}(x)$ is the best function to describe the measurement outcomes. The information about $\theta$ which can be obtained by the data $x$ is quantified by the rate of change of the likelihood function with the parameter value. A measure of such information is the zero mean value score function $\frac{\partial \ln l(\theta|x)}{\partial \theta}$. The second moment of the score is called the Fisher Information: 
\begin{equation}
F(\theta)=\int\left(\frac{\partial}{\partial\theta}\log p(x,\theta)\right)^2 p(x,\theta)dx.
\end{equation}
An important result in classical statistics is the Cram\'er-Rao bound, which gives a lower bound on the variance of $\hat{\theta}$:
\begin{equation}
V(p_{\theta},\hat{\theta})\geq\frac{1}{nF(\theta)},
\end{equation}
for $n$ repetitions of the measurement.  Hence, the Fisher information is a key figure of merit of a parameter estimation protocol. We observe that, under the assumptions of single-parameter unbiased estimation, the best estimator $\hat{\theta}_{\text{best}}$ saturates the bound.\\ 
Let us now discuss the quantum case. The state of the system under study is represented by a parametrised density matrix $\rho_{\theta}$. Let us assume that the parameter represents the information about a unitary perturbation $\rho_{\theta}=U_{\theta} \rho_0 U^\dagger_{\theta}, U_{\theta}=e^{-i H \theta}$ (Fig. \ref{fig2}). An estimator is built up by a generalised positive-operator valued measurement (POVM) $\{\Pi_x\}$ on the output state $\rho_{\theta}$, where the $\Pi_x$  denote the operators corresponding to the measurement outcomes $x$, thus obtaining $p_{\theta}(x)=\tr[\rho_\theta\Pi_x]$. The expression of the Fisher information for an arbitrary POVM is
\begin{eqnarray}\label{classf}
F(\rho_{\theta}):=\int dx \frac{1}{\tr[{\rho_{\theta} \Pi_x]}}\left(\tr[\partial_\theta\rho_{\theta} \Pi_x]\right)^2.
\end{eqnarray} 
However, the quantum scenario implies a further optimisation of the measurement \cite{helstrom,caves}. One can prove that the optimal estimator is given by a projective measurement onto the eigenbasis of the symmetric logarithmic derivative (SLD) $L$, defined implicitly as $
\frac{\partial}{\partial\theta}\rho_\theta=\frac{1}{2}(\rho_\theta L+L\rho_\theta).$ In particular, an upper bound is obtained: $F(\rho_{\theta})\leq \tr[\rho_{\theta}L^2]$. The quantum Fisher information (QFI, from now on) is then given by the optimal measurement strategy:
\begin{equation}
{\cal F}(\rho,H):=\tr[\rho L^2],
\end{equation}
where we dropped the parameter label as the QFI is independent of its value. The quantum extension of the Cram\'er-Rao bound  reads:
 \begin{eqnarray}
 V(\rho,\hat{\theta}) \geq 1/[n{\cal F}(\rho,H)],
 \end{eqnarray}
 which like the classical case is saturated asymptotically by the best estimator. The QFI enjoys a peculiar compact expression: 
 \begin{equation}
\label{fisher}
{\cal F}(\rho,H)=4\sum_{k<l}\frac{(\lambda_k-\lambda_l)^2}{\lambda_k+\lambda_l}|\bra{k}H\ket{l}|^2.
\end{equation}
where we have used the eigendecomposition of the state, $\rho=\sum_k\lambda_k\ket{k}\bra{k}$. The formula highlights that the sensitivity of a probe, and therefore its usefulness for phase estimation, is quantified by the non-commutativity of its state with the Hamiltonian. In fact, the QFI measures the sensitivity of the state $\rho$ to the unitary evolution $e^{-iH\theta}$, or, in other words, the speed of evolution of the probe under such dynamics. If and only if $H$ is diagonal in the eigenbasis of $\rho$, the transformation leaves $\rho$ invariant. It is easy to see that in that case ${\cal F}(\rho,H)=0$.
\subsubsection{Properties of the quantum Fisher information}
\label{fisherproperties}
Finally, we mention a non-exhaustive list of properties of the QFI, which will be useful in proofs later in this Section.

\begin{enumerate}
\item Up to a constant factor, the QFI is upper bounded by the variance, ${\cal F}(\rho,H)\leq 4 V(\rho,H)$, where the equality is reached for  pure states. More precisely, the QFI is the variance convex roof, ${\cal F}(\sum_i p_i \ket{\psi_i},H)=4\inf\limits_{\{p_i,\ket{\psi_i}\}} \sum_i p_i V(\ket{\psi_i},H)$ \cite{toth}.
\item The QFI is convex: ${\cal F}(p\rho_1+(1-p)\rho_2,H)\leq p{\cal F}(\rho_1,H)+(1-p){\cal F}(\rho_2,H)$, for $p$ independent of $\theta$.
\item For unitaries $U$, ${\cal F}(U\rho U^\dagger,H)={\cal F}(\rho,U^\dagger HU)$.
\item The QFI is non-increasing under CPTP maps $\Phi$ which do not depend on the parameter: ${\cal F}(\Phi(\rho),H)\leq {\cal F}(\rho,H)$.
\end{enumerate}

\subsection{Interferometry and non-classical correlations}\label{intpower}
\begin{figure}
\centering
 \includegraphics[height=6.6cm,width=12cm]{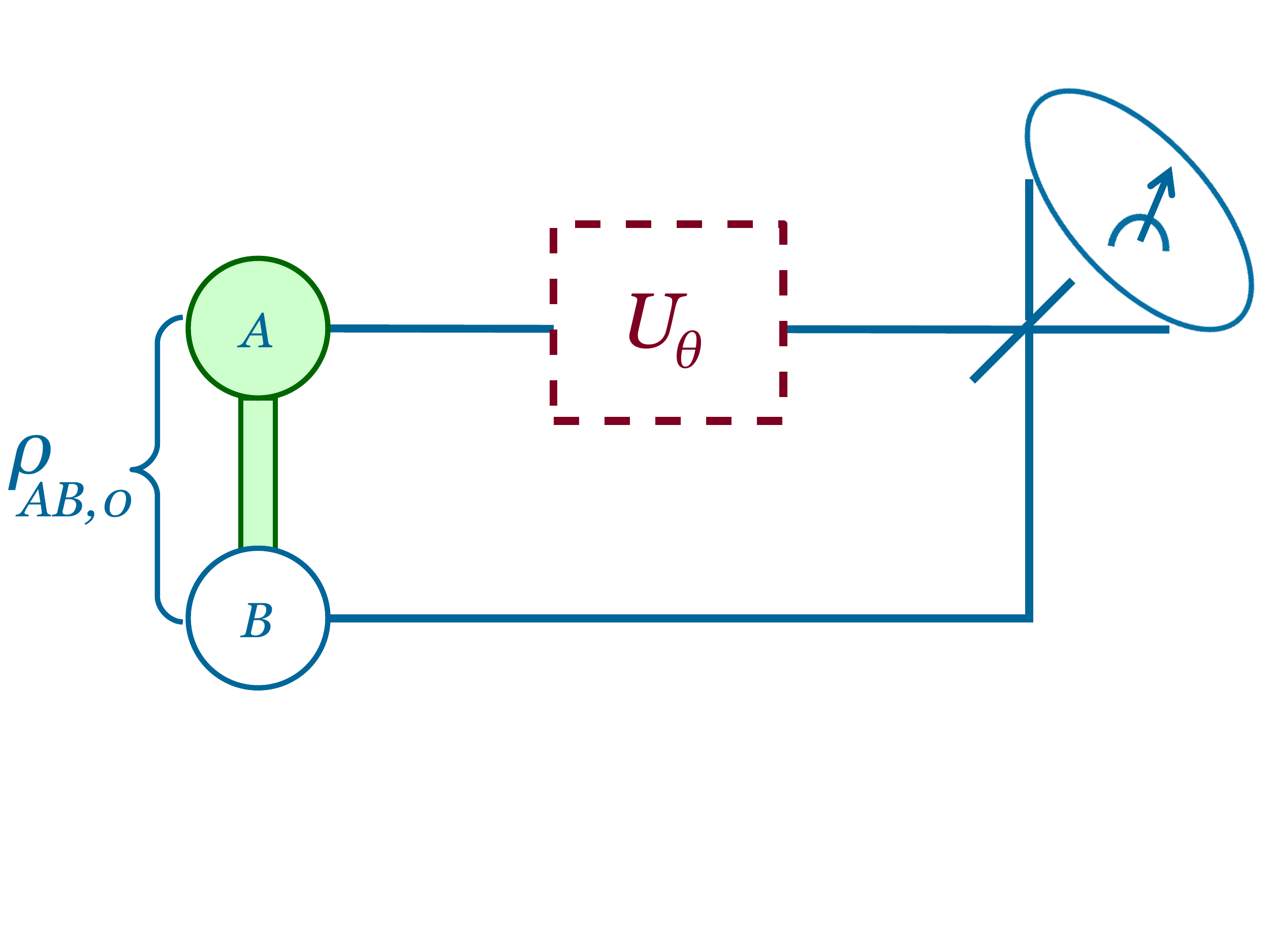}
\caption{Non-classical correlations guarantee non-vanishing precision in interferometric phase estimation. A bipartite system is prepared in an input state $\rho_{AB,0}$, and it is injected into a two-arm interferometer. A unitary transformation ${U_{\theta}}$  is applied  to subsystem $A$. The Hamiltonian eigenbasis, i.e. the phase direction, is just revealed after the interaction. The value of the imprinted phase is estimated by a measurement at the output. 
The minimum precision of the estimation, as quantified by the Interferometric Power (IP), is a measure of non-classical correlations in the input. That is, non-classical correlations ensure non-vanishing precision for any Hamiltonian.}
\label{int}
\end{figure}

An important phase estimation scenario is represented by estimation through interferometric measurements (Fig. \ref{int}). That template has been the testbed of the first observations of quantum phenomena, and it is still the standard textbook example to introduce students to quantum laws. Apart from the historical and pedagogical value, interferometry plays a premier role in modern quantum sensing schemes \cite{metrorev}. 
The architecture of an interferometric measurement is extremely simple.  A bipartite system $AB$ in the input state $\rho_{AB,0}$ is injected into a two-arm channel. Subsystem $A$ undergoes a phase shift $U_A=e^{-i  H^{\Lambda}_A \theta}$, generated by a Hamiltonian with non-degenerate spectrum $\Lambda$. This restriction is useful for understanding the role of non-classical correlations in this scenario. We remind that the phase $\theta$ represents the unknown perturbation we want to estimate, being not directly measurable. Its value is a function of the output visibility, i.e. the outcome statistics of a polarisation measurement into the output $\rho_{AB,\theta}=(U^\Lambda_A\otimes\mathbb{I}_B)\rho_{AB,0}(U^\Lambda_A\otimes\mathbb{I}_B)^\dagger$. \\
We here focus on the optimisation of the input. If the   Hamiltonian $H^{\Lambda}_A$  is fully known, then coherence of the reduced state $\rho_A$ in its eigenbasis, also called asymmetry in literature \cite{asym}, is the necessary and sufficient resource of the phase estimation. In fact, the QFI ${\cal F}(\rho,H)$ is a measure of asymmetry of the state with respect to a unitary transformation generated by $H$ \cite{yadin}. Here correlations seem not to play any role, a single party estimation is sufficient and the interferometric configuration appears redundant.
Let us now introduce a further difficulty. We suppose that the estimation is blind, in the sense that only the spectrum of the Hamiltonian generating the phase imprinting is known during the input preparation.  There is no prior information about the Hamiltonian eigenbasis. We allow to disclose the phase direction at the output, so that the measurement step can still be optimised, and the best estimator is reached. It is easy to see that there is no possible single system input $\rho_A$ guaranteeing an arbitrary degree of precision for every possible Hamiltonian. In other words, the estimation by a single party relies on pure luck as the key information about the phase direction is missing. Let us consider what happens if instead we implement the interferometer to perform the estimation. One can prove that a classically correlated  probe $AB$, or even a CQ state, are still insufficient to ensure precision for any Hamiltonian. On the other hand, by employing non-classically correlated states one can overcome the lack of knowledge about the phase direction \cite{intpow}. Similarly to what happens for the LQU, it is possible to show that a quantifier for the worst-case precision is a bona fide measure of non-classical correlations.  The optimal estimator is the one that saturates the Cram\'er-Rao bound in the limit of very large $n$, and in that case the quality of the input is determined by the QFI. The worst-case QFI  for a given state reads
\begin{equation}\mathcal{P}^{\Lambda}_A(\rho_{AB}):=\frac{1}{4}\min_{H_A}{\cal F}(\rho_{AB},H^\Lambda_A),\end{equation}
where the minimisation is over all Hamiltonians with the given non-degenerate spectrum $\Lambda$ (and where the factor 1/4 is chosen such that it cancels out the one in Eq. \ref{fisher} for the QFI under unitary dynamics).
This quantity is called   Interferometric Power (IP) of the state $\rho_{AB}$ \cite{intpow}. It quantifies the minimum sensitivity in interferometric phase estimation.

\subsubsection{Interferometric Power as a discord-like quantity}\label{proofsF}
 One can prove that the IP enjoys the same properties of the   measures of non-classical correlations, as discussed in Sec. \ref{proofs} for the LQU. We use the fact that $\mathcal{U}_A^\Lambda(\rho_{AB})\leq\mathcal{P}_A^\Lambda(\rho_{AB})$, see Sec. \ref{interplay}.
\begin{itemize}

\item[1] The IP is zero if and only if $\rho_{AB}$ is CQ. If $\rho_{AB}$ is CQ, then one can choose a Hamiltonian $H_A^\Lambda$ which is diagonal in the local basis of $A$ so that the QFI and hence the IP vanish. If, on the other hand, the IP is zero, the LQU has to be zero as well. Then we use the fact that the LQU vanishes if and only if $\rho_{AB}$ is CQ.

\item[2] The IP is invariant under local unitary transformations. It is clear from the expression in Eq. \ref{fisher} that the QFI for Hamiltonians on $A$ is invariant under local unitaries on $B$. On the same hand, local unitaries on $A$ are absorbed in the definition of $H_A^\Lambda$, thus they do not affect the minimisation.

\item[3] The IP is contractive under CPTP maps on the non-affected party $B$. This is easy to prove from the properties of the QFI itself, see \ref{fisherproperties}, or alternatively by the following, more intuitive proof. Since any map $\Phi_B$ acting on $B$  commutes with $H_A^\Lambda$, it can be included in the measurement process. Next, we note that the QFI quantifies the maximum precision that is achievable by picking the optimal estimation strategy. Since this maximum precision can only decrease when applying an extra map on $B$, we have that ${\cal F}(\rho_{AB},H_A^\Lambda)\geq {\cal F}((\mathds{1}_A\otimes\Phi_B)\rho_{AB},H_A^\Lambda)$.
\item[4] The IP reduces to an entanglement monotone for pure states. For pure states, the QFI is proportional to the variance of $H_A^\Lambda$, and the IP becomes equal to the LQU. The latter is known to be an entanglement for pure states.
\end{itemize}
\subsubsection{Restriction to $\mathds{C}^2\otimes\mathds{C}^d$}
We report  a simplified formula for the IP in  the case $A$ is a qubit, making it a computable measure of non-classical correlations for qubit-qudit systems.  
From the definition of QFI, one has ${\cal F}(\rho_{AB},aH_A^\Lambda+b\mathbb{I}_A)=a^2{\cal F}(\rho_{AB},H_A^\Lambda)$. By setting the spectrum $\Lambda$ to be $\{1,-1\}$ one has  $H_A=\vt{n}\cdot\vt{\sigma}$. The IP then becomes the minimisation of a quadratic form over the unit sphere, which leads to the following expression (like the one for the LQU of a qubit-qudit system):
\begin{equation}\mathcal{P}^{\Lambda}_A(\rho_{AB})=\lambda_{\mathrm{min}}(M_{AB}).\end{equation}
So the IP is the minimal eigenvalue of the 3$\times$3-matrix $M_{AB}$ with the following elements:
\begin{equation}(M_{AB})_{mn}=\frac{1}{2}\sum_{i,j:p_i+p_j\neq0}\frac{(p_i-p_j)^2}{p_i+p_j}\bra{\psi_i}\sigma_{mA}\otimes\mathds{1}_B\ket{\psi_j}_{AB}\bra{\psi_j}\sigma_{nA}\otimes\mathds{1}_B\ket{\psi_i}_{AB},\end{equation}
where again we have used the eigendecomposition of the state $\rho_{AB}=\sum_ip_i\ket{\psi_i}\bra{\psi_i}_{AB}$.

\subsubsection{Interplay between LQU and IP}\label{interplay}
We have shown how the LQU characterises the minimum quantum uncertainty obtained upon measuring local observables. We here point out that the skew information and the QFI, and therefore the LQU and the IP, are closely related quantities. Both the skew information ${\cal I}(\rho, H)$ and the QFI given by ${\cal F}(\rho, H)$ measure the speed of evolution of a quantum state undergoing unitary dynamics $e^{-i H \theta}$. In particular, they are associated with two  metrics included in the Fisher metrics family, which is proven to be the only class of  Riemannian metrics in the space of quantum states which is contractive under noisy maps \cite{bengtsson}. For classical probability distributions and stochastic processes, they all reduce to the classical Fisher information given in Eq. \ref{classf}.\\
We observe that the following chain of inequalities holds \cite{luo3}:
 \begin{equation}\mathcal{I}(\rho,H)\leq\frac{1}{4}{\cal F}(\rho,H)\leq 2\mathcal{I}(\rho ,H ), \forall \rho,H. \end{equation}
This implies $\mathcal{U}_A^\Lambda(\rho_{AB})\leq\mathcal{P}_A^\Lambda(\rho_{AB})$, and makes it possible to give a metrological interpretation to the LQU as well, by deriving  an upper bound for the minimum   variance in the interferometric scheme presented in Fig \ref{int}.
   In order to estimate the parameter $\theta$, we can optimise  the input state $\rho_{AB}$, the Hamiltonian $H_A$, and the final measurement. As mentioned before,  the quantum Cram\'er-Rao bound is saturated asymptotically by employing the most informative measurement, $V(\rho,\hat{\theta}_{\text{best}})=1/(n{\cal F}(\rho,H))$  \cite{metrorev,toth}. Therefore, a few algebra steps show that for interferometric phase estimations one has   \begin{equation}V(\rho_{AB},\hat{\theta}_{\text{best}})=\frac{1}{n{\cal F}(\rho_{AB},H_A^{\Lambda})}\leq\frac{1}{n \mathcal{P}^{\Lambda}_A(\rho_{AB})}\leq\frac{1}{4n\mathcal{U}^\Lambda_A(\rho_{AB})}.
\end{equation}
Hence, (the inverse of) non-classical correlations upper bound the smallest possible variance of the estimator. In other words, it is guaranteed that there exist a Hamiltonian and a measurement such that the parameter $\theta$ can be estimated with a variance lower than a value determined by the amount of discord-like correlations and the number of the experiment repetitions. Note that in this set-up we assume perfect unitary evolution and ideal measurements, but that we allow for noise in the prepared input state $\rho_{AB}$.

\subsection{Discriminating Strength}\label{ds}

We here discuss a third measure of non-classical correlations which represents the worst-case precision in another metrology task, state discrimination \cite{disc}. We also show how it relates to the LQU and therefore the IP.  \\
Suppose that we want to establish whether $n$ copies of a quantum system are prepared in a state $\rho_1$ or $\rho_2$, where each occurs with equal probability. It is allowed to obtain information by measuring the system. According to the Holevo-Helstrom theorem, the minimum error probability after optimising over all possible POVMs is given by
\begin{equation}P^{(n)}_\mathrm{err,min}:=\frac{1}{2}\left(1-\frac{1}{2}||\rho_1^{\otimes n}-\rho_2^{\otimes n}||_1\right),\end{equation}
where the optimal POVM discriminates the positive and negative eigenspaces of $\rho_1^{\otimes n}-\rho_2^{\otimes n}$. In the asymptotic limit of large $n$, the minimum error probability follows an exponential decay law
\begin{equation}P^{(n)}_\mathrm{err,min}\approx e^{-n\xi(\rho_1,\rho_2)},\end{equation}
where the decay constant is given by
\begin{equation}\xi(\rho_1,\rho_2):=-\lim_{n\rightarrow\infty}\frac{\ln P^{(n)}_\mathrm{err,min}}{n}=-\ln\left(\min_{0\leq s \leq 1} \tr[\rho_1^s\rho_2^{1-s})\right].\end{equation}
Such a limit is called quantum Chernoff bound \cite{chernov}.
Finally, we define the quantity
\begin{equation}Q(\rho_1,\rho_2):=e^{-\xi(\rho_1,\rho_2)}=\min_{0\leq s \leq 1} \tr[\rho_1^s\rho_2^{1-s}].\end{equation}
It is immediately clear that $0\leq Q(\rho_1,\rho_2)\leq\tr[\rho_1^{1/2}\rho_2^{1/2}]\leq 1$, and, if at least one of the two states is pure, $Q(\rho_1,\rho_2)$ reduces to Uhlmann's fidelity $F(\rho_1,\rho_2):=\left(\tr[\sqrt{\sqrt{\rho_1}\rho_2\sqrt{\rho_1}}]\right)^2$.

A state discrimination problem represents the discretised version of a phase estimation scenario, where instead of a continuous parameter $\theta$ one wishes to know the value of a two-value label identifying one of the two options $\rho_{1,2}$. It is then not surprising that non-classical correlations play a role in an interferometric state discrimination scheme called quantum illumination \cite{lloyd,pira}. The protocol runs as follows. An experimentalist Alice prepares $n$ copies of a bipartite state $\rho_{AB}$, where $A$ is the probe part and $B$ is a reference system. A second player Charlie chooses an undisclosed unitary  $C_A$ from a given set of allowed transformations $\mathcal{S}$. Then, Alice sends her $n$ copies to Charlie who is free to   either  leave the $n$ copies unaltered, or rotate all of them by implementing $C_A$. Finally, Alice has to decide which of the two actions Charlie has chosen, being allowed to perform any POVM on the $n$ copies. That means that she has to  discriminate between $\rho_1^{\otimes n}=\rho_{AB}^{\otimes n}$ and $\rho_2^{\otimes n}=(C_A \rho_{AB} C_A^\dagger)^{\otimes n}$. The Discriminating Strength (DS) of the probe state $\rho_{AB}$ is defined as the Alice discriminating ability in the worst possible case:
\begin{equation}\mathcal{D}_{A}^{\mathcal{S}}(\rho_{AB}):=1-\max_{C_A\in \mathcal{S}}Q\left(\rho_{AB},C_A\rho_{AB} C_A^\dagger\right).\end{equation}
From the definition of the quantum Chernoff bound, it is clear that $A$ is able to perform better if the DS is higher. Note that this is again a context in which there is a clear asymmetry between the role played by the parts of a bipartite system. 

So far, we have not specified what the set of allowed transformations $\mathcal{S}$ is, and the DS of course depends heavily on the choice of this set. A first observation is that if $\mathcal{S}$ were chosen to be the whole group of unitaries on $A$, the DS would always be zero as this group includes the identity. Clearly, we need to avoid such pathological cases. We restrict Charlie's choice to the set of unitaries $C_A=\exp(iH_A^\Lambda)$. In this parametrisation, $H_A^\Lambda$ is a Hamiltonian acting on $A$, with non-degenerate spectrum $\Lambda$ (notice the similarity with the LQU case): $H_A^\Lambda=U_A\text{diag}(\Lambda) U_A^\dagger,$ where $U_A\in \mathrm{U}(d_A)$. The DS is then defined as
\begin{equation}\mathcal{D}_{A}^\Lambda(\rho_{AB}):=1-\max_{H_A^\Lambda}Q\left(\rho_{AB},e^{iH_A^\Lambda}\rho_{AB} e^{-iH_A^\Lambda}\right).\end{equation}
A crucial point to discuss is  to what extent the DS depends on the choice of the spectrum $\Lambda$. Although in Ref.\cite{disc} the authors mention that it is tempting to conjecture that the harmonic spectrum (i.e. $\lambda_i-\lambda_{i+1}$ is constant, $\forall i$) should be optimal, no clear answer to this question is given. One obvious property of the DS is the invariance under constant shifts, $\mathcal{D}_{A}^\Lambda(\rho)=\mathcal{D}_{A}^{\Lambda+b}(\rho)$, $\forall \rho$, $b\in\mathds{R}$.

\subsubsection{Discriminating Strength as a measure of non-classical correlations}
The intuition behind the DS is that establishing whether the state has undergone a local rotation should be easier the more the part $A$ potentially affected by the rotation  is non-classically correlated with an unaffected part $B$. We here report the proof that the DS is a bona fide measure for non-classical correlations, as it has the same properties of the LQU and the IP discussed in Secs. \ref{proofs} and \ref{proofsF} respectively.
\begin{itemize}
\item[1] The DS is zero if and only if $\rho$ is CQ. The DS is zero if and only if there is a $C_A$ such that $Q(\rho,C_A\rho C_A^\dagger)=1$, which is the case if and only if $\rho=C_A\rho C_A^\dagger$. Since $C_A$ has a non-degenerate spectrum, this is equivalent to requiring that $\rho$ and $H_A^\Lambda$ are diagonal in the same basis, i.e. $\rho$ is CQ.

\item[2] The DS is invariant under local unitary transformations. First we note that $(U\rho U^\dagger)^s=U\rho^sU^\dagger$ for any unitary $U$. Using this property and the cyclicity of the trace, it follows that $Q$ is invariant under local unitaries on $B$. For local unitaries on $A$, we can use the same property and  absorb the transformation in the Hamiltonian, since maximising over $U_A^\dagger H_A^\Lambda U_A$ is equivalent to maximising over $H_A^\Lambda$.

\item[3] The DS is contractive under CPTP maps on the unchanged party $B$. Since any local map   $\Phi_B$ commutes with the transformation on $A$ induced by $H_A^\Lambda$, $\Phi_B$ can be absorbed in the POVM. The minimum error probability is obtained by minimising the error probability over all POVMs on $\rho_{AB}^{\otimes n}$. Absorbing the extra local map $\Phi_B$ can only increase the error probability, and hence $Q$ is monotonically increasing: $\mathcal{D}^\Lambda_A(\Phi_B(\rho_{AB}))\leq\mathcal{D}^\Lambda_A(\rho_{AB})$.

\item[4] The DS reduces to an entanglement monotone for pure states. If $\ket{\psi}_{AB}$ is transformed to $\ket{\phi}_{AB}$ under LOCC operations, we can write \begin{eqnarray}\ket{\phi}\bra{\phi}_{AB}=\sum_iM_{i,A}V_{i,B}\ket{\psi}\bra{\psi}_{AB}M^\dagger_{i,A}V^\dagger_{i,B},\end{eqnarray}
 where $\{M_{i,A}\}$ are Kraus operators on $A$ and $\{V_{i,B}\}$ are unitaries on $B$. One has $M_{i,A}V_{i,B}\ket{\psi}_{AB}=\sqrt{p_i}\ket{\phi}_{AB}$. Similarly to the case of the LQU, one can prove that maximising over Hamiltonians on $A$ is equivalent to maximising over Hamiltonians on $B$. Assume that $\tilde{H}^\Lambda_B$ achieves that maximum. Then one obtains 
\begin{eqnarray}
D^\Lambda_A(\ket{\phi}\bra{\phi}_{AB})&=&1-\max_{H^\Lambda_B}\left|\bra{\phi}e^{iH^\Lambda_B}\ket{\phi}_{AB}\right|^2=1-\sum_i\frac{1}{p_i}\max_{H^\Lambda_B}\left|\bra{\psi}M^\dagger_{i,A}e^{iH^\Lambda_B}M_{i,A}\ket{\psi}_{AB}\right|^2 \nonumber \\
&\leq&1-\sum_i\frac{1}{p_i}\left|\bra{\psi}M^\dagger_{i,A}e^{i\tilde{H}^\Lambda_B}M_{i,A}\ket{\psi}_{AB}\right|^2\nonumber\\
&\leq&1-\left|\bra{\psi}\sum_iM^\dagger_{i,A}M_{i,A}e^{iH^\Lambda_B}\ket{\psi}_{AB}\right|^2\nonumber\\
&=&1-\left|\bra{\psi}e^{i\tilde{H}^\Lambda_B}\ket{\psi}_{AB}\right|^2=D^\Lambda_A(\ket{\psi}\bra{\psi}_{AB}).
\end{eqnarray}
Note that in the first line $V_{i,B}$ and $V_{i,B}^\dagger$ are included into the maximisation over $H_B^\Lambda$; in the second line, we rely on the fact that the maximum of a function is lower bounded by the function evaluated at any given point, and the Cauchy-Schwarz inequality.
\end{itemize}

\subsubsection{Interplay with the Local Quantum Uncertainty}
The DS is related to the LQU. To show that, we remind that, for any given density matrix $\rho$ and Hermitian operator $O$, the following result holds:
\begin{equation}\label{property}\min_{0\leq s \leq1}\tr[\rho^s O\rho^{1-s}O]=\tr[\rho^{1/2}O\rho^{1/2}O].\end{equation}
This is clear by writing $\rho$ in terms of its eigenvectors $\{\ket{\psi_i}\}$ and by employing a non-increasing order for the eigenvalues $\lambda_i$:
\begin{equation}\min_{0\leq s \leq1}\tr[\rho^s O\rho^{1-s}O]=\sum_i \lambda_i |\bra{\psi_i}O\ket{\psi_i}|^2+\min_{0\leq s\leq1}\sum_{i<i'}(\lambda_i^s\lambda_{i'}^{1-s}+\lambda_{i'}^s\lambda_{i}^{1-s})|\bra{\psi_i}O\ket{\psi_i'}|^2.\end{equation}
It is then easy to see that for each term in the second sum  the minimum is achieved for $s=1/2$, which proves the  result.
The link between LQU and DS  is manifest by Taylor expanding $e^{iH^\Lambda_A}$ with respect to $\Lambda$:
\begin{eqnarray}\mathcal{D}_A^\Lambda(\rho_{AB})&=&1-\max_{\{H_A^\Lambda\}}\min_{0\leq s \leq 1}\tr[\rho_{AB}^se^{iH^\Lambda_A}\rho_{AB}^{1-s}e^{-iH^\Lambda_A}]\nonumber\\
&=&-\max_{\{H_A^\Lambda\}}\min_{0\leq s \leq 1}\tr[\rho_{AB}^sH_A^\Lambda\rho_{AB}^{1-s}H_A^\Lambda-H_A^\Lambda\rho_{AB} H_A^\Lambda]+O(\Lambda^3)\nonumber\\
&=&-\max_{\{H_A^\Lambda\}}\tr[\rho_{AB}^{1/2}H_A^\Lambda\rho_{AB}^{1/2}H_A^\Lambda-H_A^\Lambda\rho_{AB} H_A^\Lambda]+O(\Lambda^3)\nonumber\\
&=&\min_{\{H_A^\Lambda\}}\tr[H_A^\Lambda\rho_{AB} H_A^\Lambda-\rho_{AB}^{1/2}H_A^\Lambda\rho_{AB}^{1/2}H_A^\Lambda]+O(\Lambda^3)\nonumber\\
&=&\mathcal{U}^\Lambda_A(\rho_{AB})+O(\Lambda^3).
\end{eqnarray}
where we have used Eq. \ref{property} in the third line.

We observe that  for small $\Lambda$ the LQU can be interpreted as the DS in a discrimination task. In this statement,  small $\Lambda$ means that the local transformations should be close to the identity, i.e. only small perturbations are allowed.

\subsubsection{Computable expressions of the DS}

In Ref. \cite{disc}, the authors present expressions for the DS in a few special cases. We only give details about the derivation of the formula for qubit-qudit states, as done before for the LQU and IP, but other cases are mentioned for the sake of completeness.

First, let us consider pure bipartite states $\ket{\psi}_{AB}$. The Schmidt decomposition is given by $\ket{\psi}_{AB}=\sum_{i=1}^{\min\{d_A,d_B\}}\sqrt{\sigma_i}\ket{i}_A\ket{i}_B$ with Schmidt coefficients $\{\sigma_i\}$. Then, the DS is given by the following expression:
\begin{equation}
\mathcal{D}^\Lambda_A(\ket{\psi}\bra{\psi}_{AB})=1-\max_{\pi_\alpha}\left|\sum_k\sigma_{\pi_\alpha[k]}e^{i\lambda_k}\right|^2.
\end{equation}
where we now have a maximisation over the group of permutations $\pi_\alpha$ on the Schmidt coefficients $\{\sigma_i\}$, instead of the maximisation over all Hamiltonians $H_A^\Lambda$ with spectrum $\Lambda$ (which is an infinite set). If $d_A>d_B$, the set of Schmidt coefficients should be extended with zeros to obtain a set of size $d_A$.

We mentioned before that it is tempting to hypothesise that the DS obtained by fixing an harmonic spectrum would yield the most accurate measure for non-classical correlations. Even though it is not clear if this is true, it explains why it is interesting to calculate  the expression of the DS in such a case. By defining the  fundamental frequency $\omega:=|\lambda_i-\lambda_{i+1}|\leq2\pi/d_A$, 
we can further simplify the previous formula. The permutation which maximises the second term  gives the following values: $\sigma_1=0$, $\sigma_2=\omega$, $\sigma_3=-\omega$, $\sigma_4=2\omega$, $\sigma_5=-2\omega$, et cetera. The resulting expression for the DS is then
\begin{equation}
\mathcal{D}^\Lambda_A(\ket{\psi}\bra{\psi}_{AB})=1-\left|\sum_{n=0}^{[(d_A+1)/2]-1}\sigma_{2n+1}e^{in\omega}+\sum_{n=1}^{d_A-[(d_A+1)/2]}\sigma_{2n}e^{in\omega}\right|^2.
\end{equation} 
The precise details of this expression are not very relevant to our discussion, but it is noteworthy that we have managed to get rid of the optimisation over the unitaries.

We now analyse the qubit-qudit case (where the subsystem $A$ is the qubit). The DS is invariant under constant shifts, so we can parametrise the spectrum as $\{-\lambda,\lambda\}$. The Hamiltonian takes the form $H_A^\Lambda=\lambda\vt{n}\cdot\vt{\sigma}_A$. For conciseness of notation, we introduce $\vt{\sigma}_{A,n}:=\vt{n}\cdot\vt{\sigma}_A$. The quantum Chernoff bound then reads
\begin{eqnarray}
Q(\rho_1,\rho_2)&=&\min_{0\leq s\leq1}\tr\left[\rho_{AB}^se^{i\lambda\vt{\sigma}_{A,n}}\rho_{AB}^{1-s}e^{-i\lambda\vt{\sigma}_{A,n}}\right]\nonumber\\
&=&\cos^2\lambda+\min_{0\leq s\leq1}\tr[\rho_{AB}^s\vt{\sigma}_{A,n}\rho_{AB}^{1-s}\vt{\sigma}_{A,n}]\sin^2\lambda\nonumber\\
&=&\cos^2\lambda+\tr[\rho_{AB}^{1/2}\vt{\sigma}_{A,n}\rho_{AB}^{1/2}\vt{\sigma}_{A,n}]\sin^2\lambda.
\end{eqnarray}
Using this expression, we finally get the formula
\begin{eqnarray}
\mathcal{D}^\Lambda_A(\rho_{AB})&=&\min_{\vt{n}}\left(1-\tr[\rho_{AB}^{1/2}\vt{\sigma}_{A,n}\rho_{AB}^{1/2}\vt{\sigma}_{A,n}]\right)\sin^2\lambda\nonumber\\
&=&\mathcal{U}_A^\Lambda(\rho_{AB})\frac{\sin^2\lambda}{\lambda^2}.
\end{eqnarray}
To summarise, there is a proportionality relation between the DS and the LQU for qubit-qudit systems, which turns out to be an equality when $\lambda$ approaches zero, as $\frac{\sin^2\lambda}{\lambda^2}\rightarrow 1$.

\section{Conclusion}
We here reviewed recent works providing a metrological interpretation to non-classical correlations. Our understanding of an elusive, information-theoretic concept has been shaped by linking it to experimentally testable effects. State-observable complementarity implies genuine quantum uncertainty. Such uncertainty corresponds to sensitivity to a quantum evolution. The state rate of change triggers measurement precision of a complementary property. The peculiar asymmetry of non-classical correlations finds an operational interpretation in metrology, when such an argument is extended to compound systems.  If and only if the state of a bipartite system shows non-classical correlations, Quantum Mechanics dictates sensitivity to local perturbations,  which translates into a guaranteed minimum performance in  paradigmatic scenarios such as parameter estimation and state discrimination. The LQU, the IP and the DS are parent discord-like measures which
capture this distinctive feature of quantum states. An interesting question is to establish if the metrologic measures of discord, which have been introduced to catch
bipartite statistical dependence, can be extended to quantify multipartite correlations. We are actively working  on the problem and we are able to anticipate that the answer is positive, while a complete study on the topic will be published in the near future.   Such extension relies on employing   non-unitary evolutions, where the information is imprinted by noisy channels. The scenario will provide an operational interpretation of multipartite non-classical correlations in more realistic scenarios, taking in account  non-negligible errors in both state and gate preparations, and the presence of an environment. \\
Finally, we would like to point the Reader to further results on metrological measures of non-classical correlations.  An experimental comparison of classical and quantum resources in interferometric phase estimation  has been implemented in a room temperature NMR (Nuclear Magnetic Resonance) system \cite{intpow}.  Extensions of the reported results to continuous variable systems have been obtained \cite{gaussint,gaussdisc,gauss2}. Other geometric measures of non-classical correlations inspired by metrological tasks have also been proposed \cite{sp,resp}.

\end{document}